# Reduced Dark Current in $Cd_{0.9}Zn_{0.1}Te$ Detector Arrays by Aluminium Oxide Passivation

Sudarshan Singh, Luc Montpetit, Eloïse Rahier, Mahmoud R. M. Atalla, Sebastian Koelling, Oussama Moutanabbir*

Department of Engineering Physics, École Polytechnique de Montréal, Montréal, C.P. 6079, Succ. Centre-Ville, Montréal, Québec, Canada H3C 3A7

**Abstract:** Contrary to the prevailing view that processing cadmium zinc telluride (CZT) semiconductors above 150 °C leads to irreversible device degradation, here we show that $Al_2O_3$ passivation layer deposited by atomic layer deposition at 250 °C not only preserves CZT detector performance but also substantially improves it. Pixelated metal-semiconductor (MS) and metal-insulator-semiconductor (MIS) CZT detectors were passivated with a thin atomic-layer-deposited $Al_2O_3$ film and characterized electrically. The passivated devices exhibited no measurable degradation, even under a high-bias operation of 1000 V. Instead, the dark current decreased by approximately a factor of five, while the interpixel leakage current was reduced by nearly one order of magnitude, from ~41 nA to ~2-3 nA at -200 V. Passivation also produced highly uniform dark-current characteristics across adjacent pixels and completely eliminated current-voltage hysteresis, indicating suppression of defect-assisted charge transport. Furthermore, no significant difference in dark current was observed between the MS and MIS detectors after passivation, suggesting that the $Al_2O_3$ layer dominates the surface electrical behavior. These results demonstrate that optimized $Al_2O_3$ passivation at 250 °C is fully compatible with high-performance CZT detector processing and provides a practical route toward lower-noise, higher-spectral-resolution X-ray imaging systems.

Cadmium zinc telluride (CZT) is a promising semiconductor for high-energy X-ray detection at room temperature, offering higher detection efficiency and a smaller footprint than traditional Si and Ge detectors, which have limited stopping power and require cryogenic cooling[1]. Its higher average atomic number, wider bandgap, and higher resistivity lead to improved X-ray absorption, electron-hole pair creation, and reduced noise, thereby enhancing signal quality [2]. As a result, CZT detectors are increasingly employed in gamma cameras, single-photon emission computed tomography (SPECT) systems, and bone density scanners, and their application is now being extended to computed tomography (CT)[2–4]. Notably, CZT enables photon-counting and spectral imaging, which can reduce radiation dose while enhancing image contrast [1].

Despite its remarkable potential, the widespread application of CZT technology remains constrained by the stringent requirements for high-purity, low-defect, and compositionally homogeneous semi-insulating crystals [5]. Moreover, wafer processing, including mechanical polishing and chemo-mechanical polishing (CMP), modifies the surface electrical properties and can introduce significant disparities between surface and bulk resistivity [6,7]. Consequently, effective surface passivation of CZT detectors is essential to suppress leakage currents and achieve optimal detector performance. Most passivation treatments are categorized into two processes, namely wet and dry. Despite their widespread adoption, wet passivation processes involve complex interfacial reactions at the CZT surface that are poorly understood and often suffer from limited reproducibility [8,9]. In contrast, dry passivation techniques provide greater control over film uniformity and interface quality while simultaneously offering device encapsulation. Using these methods, thin films such as $SiN_x$ [10], $Al_2O_3$ [11,12], AlN [13], and diamond-like carbon [14] have demonstrated reduced surface leakage current (SLC) and improved detector performance. Among these, $Al_2O_3$ has attracted significant attention for its high dielectric constant, excellent chemical stability, and inherent negative fixed oxide charges, which enhance field-effect passivation and reduce surface electron conduction [13,15]. Zanettini et *al*. reported reduced SLC in CZT detectors passivated with a 50-nm sputtered $Al_2O_3$ layer [11]. Subsequently, $Al_2O_3$ passivation via atomic layer deposition (ALD) showed enhanced charge-collection efficiency and carrier lifetime in CdTe detectors [13], and improved energy resolution in CZT thick films [16]. Typical $Al_2O_3$ passivation process is performed below 150-200 °C to prevent possible device or material deterioration at high temperature. However, a mixed behavior, with both improvement and degradation in the CdTe and CdZnTe devices, was observed previously following high-temperature annealing [17–22]. Moreover, the electrical quality and mechanical integrity of ALD-grown $Al_2O_3$ films typically improve with increasing deposition temperature,[23] thereby enhancing surface resistivity and long-term detector stability. Nevertheless, studies on passivation of the CZT detector with highly insulating $Al_2O_3$ film deposited at a relatively higher temperature are still conspicuously missing in the literature.

Herein, we address this very issue by investigating two sets of pixelated CZT detectors with distinct device architectures, metal-semiconductor (MS) and metal-insulator-semiconductor (MIS). The inclusion of both device configurations enables assessment of whether the benefits of $Al_2O_3$ passivation arise solely from surface modification or are further enhanced when the dielectric layer actively participates in the device structure. Prior to device fabrication, CZT samples were thoroughly characterized for their structural, surface morphology, and chemical composition. Finally, the devices were passivated with thin ALD-grown $Al_2O_3$ layers deposited at 250 °C and electrically evaluated before and after passivation. The results show that surface passivation at a higher deposition temperature significantly improves device performance, evidenced by lower dark current and trap annihilation, with no signs of device degradation.

CZT ($Cd_{0.9}Zn_{0.1}Te$) single crystals used in these investigations were grown by the Travelling Heater Method with feed polycrystalline materials with a purity higher than 6N. Prior to device processing, in-depth surface and bulk characterization studies were conducted to assess the quality of the crystals. Optical and scanning electron microscopic (SEM) images of the material surface reveal the imprint of alumina abrasives used during the polishing process (Figs. S1(a) and S1(b)). The infrared transmission image of the sample (Fig. S1(c)) indicates a higher density of Te inclusions than observed at the surface by optical microscopy, indicating that these inclusions are embedded in the bulk material[24]. These inclusions were predominantly 5-15 μm in size, with no inclusions larger than 15 μm observed. Figure S1(d) depicts a SEM image of a typical hexagonal Te inclusion on the CZT wafer surface. The atomic force microscope (AFM) measurements (Fig. S1(e)) reveal that the root-mean-square (RMS) roughness of the samples over an area of 10×10 $\mu m^2$ is ~0.6 nm, which is substantially lower than previously reported values for other surface treatments[25,26]. The high-resolution X-ray photoelectron spectroscopy (XPS) measurements (Fig. S2) show the Te 3d spectra, with peaks at 572.8 eV and 583.2 eV corresponding to the $3d_{5/2}$ and $3d_{3/2}$ levels of $Te^{2-}$, respectively, and a spin-orbit splitting of 10.39 eV. Additional peaks at higher binding energies, 576.3 eV and 586.7 eV, correspond to $Te^{4+}$, indicating the presence of native tellurium oxides [27]. The native oxide thickness was estimated to be less than 1 nm using the Beer-Lambert law for a single-photoelectron take-off angle[27]. The overall Te/CdZn composition ratio was found to be approximately 1.

The deconvolution of the O1s spectrum (Fig. S2(d)) reveals two peaks centered at 530.6 eV and 531.4 eV, corresponding to the $TeO_2$ layer and surface hydroxides, respectively, indicating that oxygen is predominantly present as $TeO_2$ [16]. Despite eliminating surface scratches and providing a smoother surface, the CMP process may also introduce sub-surface structural damage extending from the nanometre to micrometer scale [28]. Therefore, transmission electron microscopy (TEM) analyses were carried out prior to the deposition of pixel electrodes to investigate the bulk and subsurface characteristics of the CZT crystal. Figure 1(a) presents a focused-ion beam (FIB) image acquired during the preparation of a cross-section

TEM lamella using the standard lift-out procedure. A high-resolution (HR) TEM image near the CZT surface is shown in Fig. 1(b), where the bright layer at the top corresponds to a protective carbon coating deposited during sample preparation. No noticeable damage was observed near the surface. The Fast Fourier Transform (FFT) of the HR-TEM image (Fig. 1(c)) represents the reciprocal lattice of the crystal, with the crystallographic directions oriented perpendicular to the planes observed in the real-space image. The [111] direction shown in yellow was identified with a measured interplanar spacing of 3.7 Å. The [111] direction in the FFT image, being perpendicular to the surface of the sample in the HR image, confirms that the (111) planes are parallel to the surface of the wafer. Finally, the high-angle annular dark-field (HAADF) and scanning transmission electron microscope and energy dispersive spectroscopy (STEM-EDS) have been performed to probe the composition of the CZT. The elemental maps presented in Figs. (e)-1(h) show a homogeneous distribution of Cd, Zn, and Te within the sample near the surface.

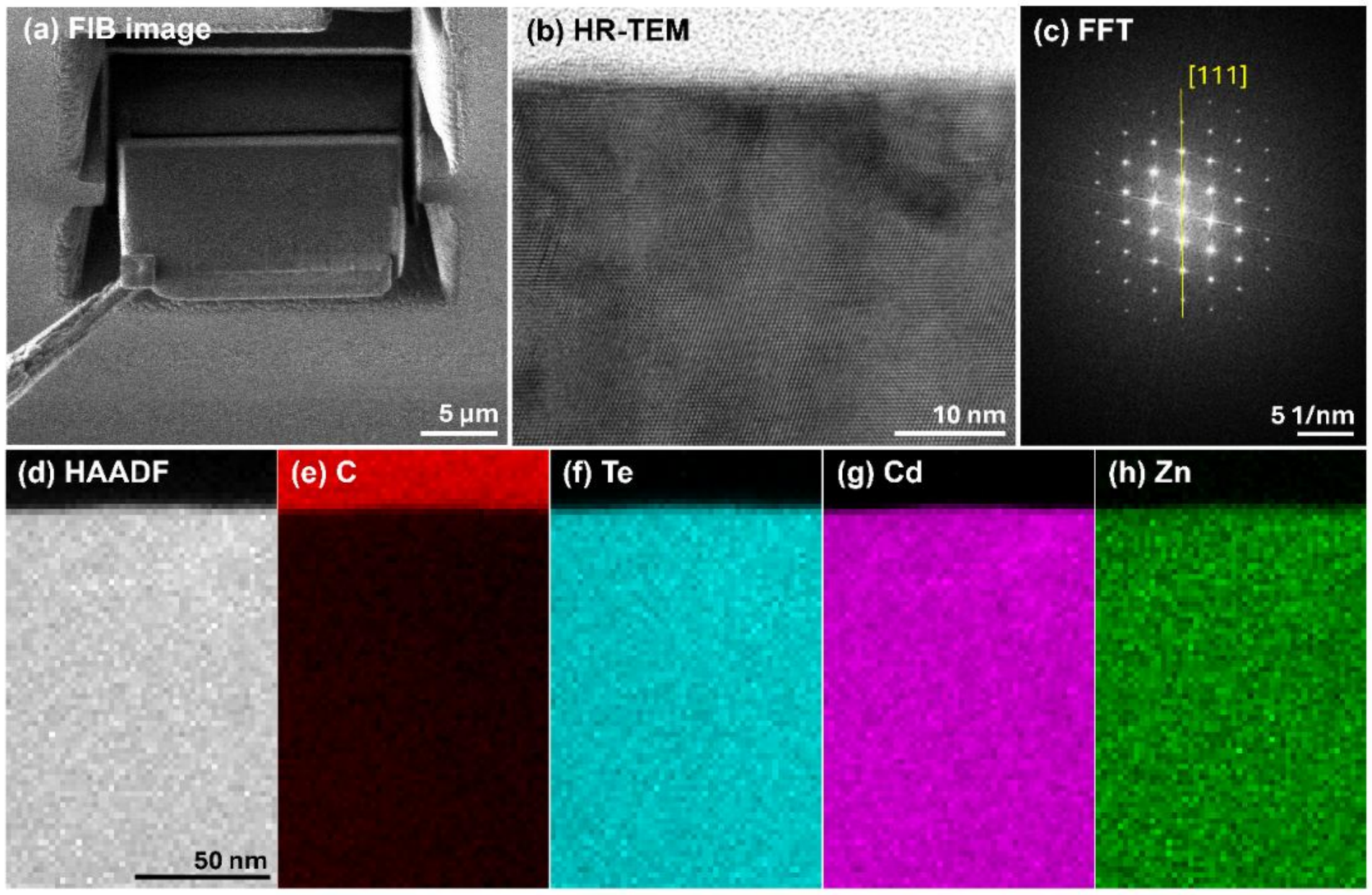

**FIG. 1:** (a) FIB image of the lift-out process of the cross-section for TEM lamella preparation. (b) High-resolution TEM of a CZT surface. (c) Fast Fourier Transform (FFT) of the HR image on the left with the [111] crystallographic direction highlighted in yellow. (d) - (h) HAADF and STEM-EDS maps of C, Te, Cd and Zn, respectively.

We fabricated detector arrays with a pixelated electrode on the B-face, whereas a planar electrode on the A-face. The detailed device fabrication process is provided in the supporting information. In the MIS device structure, a 3 nm-thick $Al_2O_3$ interlayer was introduced between the top pixelated electrode and the CZT surface. This $Al_2O_3$ film can act as a passivation layer to suppress interface recombination and can play a crucial role in controlling the device leakage current[29]. Finally, the interpixel spacing and sidewalls of both device architectures were further passivated with a 10 nm-thick $Al_2O_3$ layer. Both the 3 nm and 10 nm $Al_2O_3$ films were deposited at a substrate temperature of 250 °C. Since the electrical quality of the $Al_2O_3$ film is critical for effective interface passivation, its insulating properties were evaluated by depositing a 10 nm thick $Al_2O_3$ layer on a Si substrate. The $Al_2O_3$ films deposited at 250 °C were found to exhibit significantly lower leakage current than those deposited at lower temperature (150 °C), as shown in Fig. S3(a). Moreover, no measurable current was observed across the film in either top-top or top-bottom contact configurations at ±1 V bias for the film deposited at 250 °C (Fig. S3(b)). Therefore, the ALD process was carried out at the optimal deposition temperature of 250 °C for all devices processed in this study. A Si piece placed adjacent to the CZT samples during the ALD process was used to characterize the deposited films independently. Figure S4 shows a typical AFM image of a 10 nm thick $Al_2O_3$ layer deposited at 250 °C, revealing a smooth surface with RMS roughness of ~0.2 nm. The atomic concentrations of Al 2p and O 1s, determined by XPS (Fig. S5), were approximately 41% and 59%, respectively, consistent with the stoichiometry of $Al_2O_3$. Figures 2(a)-2(c) illustrate schematic representations of the fabricated MS, MS + passivation and MIS + passivation device structures, respectively, where the MS + passivation configuration corresponds to the same MS device after passivation. The SLC was measured and compared across these three different configurations. During the SLC measurements between adjacent pixel pairs, the bottom electrode and all non-measured top pixels were left floating.

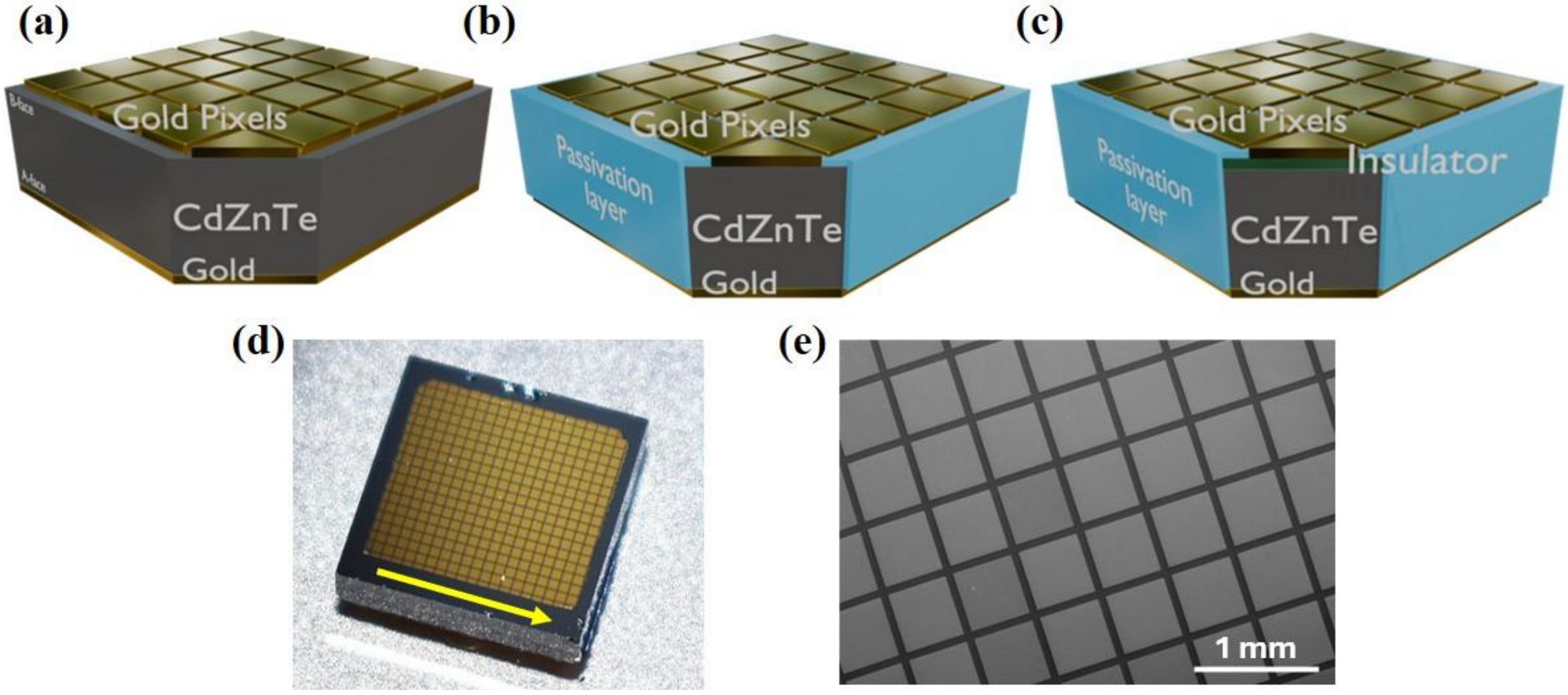

**FIG. 2:** Schematic diagram of the fabricated devices with different device configurations: (a) metal-semiconductor-metal, (b) metal-semiconductor-metal with passivation, (c) metal-insulator-semiconductor with passivation, (d) exemplary digital photo of the fabricated pixelated device and corresponding SEM image (e).

Figure 3(a) depicts the SLC values measured sequentially between adjacent pixel pairs across the detector array, along the yellow arrow indicated in Fig. 2(d). Nearly identical dark current values are observed for all pixel pairs, confirming the uniformity of the pixel fabrication process and the homogeneity of the device surface. Before passivation, the MS device exhibits an average SLC of approximately 41 nA between adjacent pixels at an applied bias of -200 V. In contrast, both $Al_2O_3$-passivated MS and MIS devices demonstrated a significant reduction in SLC by approximately a factor of 18, resulting in an average value of ~2.3 nA. Furthermore, the inter-pixel dark current was evaluated as a function of pixel-to-pixel separation, from the nearest pair (1–2) to the farthest pair (1–16), for all three device configurations and the results are shown in Fig. 3(b). As expected, the non-passivated MS device exhibits a relatively high dark current that decreases with increasing pixel separation, consistent with the corresponding increase in inter-pixel resistance. The pixelated contacts are deposited by e-beam evaporation and exhibit non-ohmic contact. Therefore, accurately estimating the surface resistivity of CZT samples is challenging.

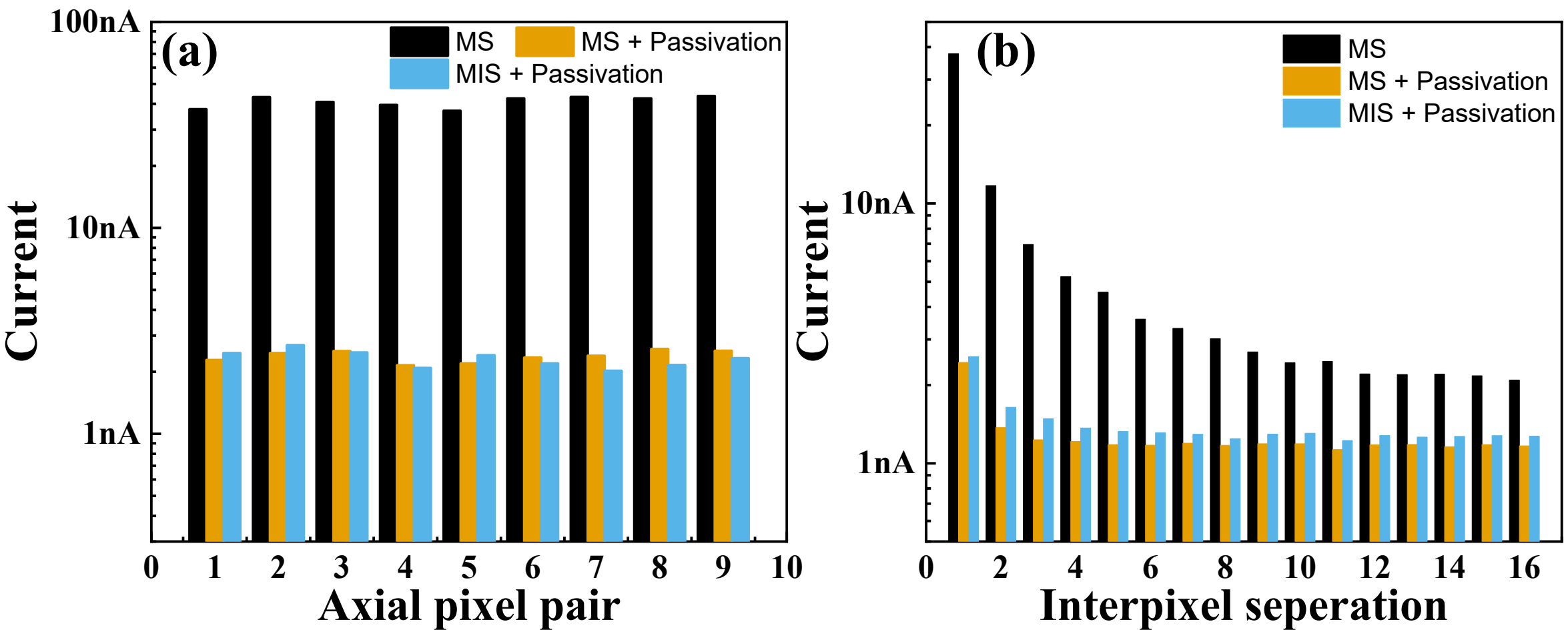

**FIG. 3:** (a) Surface leakage current of MS, MS + Passivation and MIS + Passivation devices measured between adjacent pixel pairs starting from the 1st pair on the left side to the 9th on the right side (as shown by the arrow in Fig. 2(d)). The top of the bars corresponds to the current value measured at a 200 V difference between pixels. (b) The current value measured between two pixels starts from the nearest one (pixels 1-2) and extends to the farthest (pixels 1-16).

Nevertheless, for the as-fabricated non-passivated device, the dark current decreases by approximately one order of magnitude from the nearest to the farthest pixel pair. In contrast, both MS and MIS devices with $Al_2O_3$ passivation exhibit two distinct features in the dark current. Firstly, the adjacent pixel pairs show a reduction in dark current by more than one order of magnitude. Secondly, both devices show only a slight reduction in current from ~2.5 nA to ~1.5 nA up to the fourth pixel, after which the current remains nearly constant. This indicates the possible suppression of interpixel conducting channels, leading to a significant increase in surface resistivity after passivation. Moreover, the intrinsic negative charges in the $Al_2O_3$ film can induce a field-effect depletion region in the near-surface CZT layer, thereby restoring bulk-like resistivity, suppressing surface current losses, and providing effective passivation between the pixelated electrodes [11,13,15]. These results demonstrate that a thin ALD-deposited $Al_2O_3$ layer can significantly suppress SLC in CZT detectors, compared to a 55-nm-thick sputtered $Al_2O_3$ layer [11]. Interestingly, the MS and MIS devices exhibit nearly identical dark currents, suggesting that the insertion of a thin insulating $Al_2O_3$ layer between the metal and semiconductor does not provide additional dark current suppression. This could be attributed to the dielectric breakdown of the $Al_2O_3$ film under high applied voltage.

The current-voltage (I-V) characteristics of all device configurations were measured at a higher bias voltage between the top and bottom electrodes, and their dark currents were compared. It is well known that CZT contains a high density of deep-level traps and surface states, therefore, establishing equilibrium between free and trapped charges may take several seconds to minutes[30]. To account for this, the I-V measurements were performed under two different delay times. The first is the default auto delay (1 ms), whereas the second has a source delay of 500 ms. Figure 4(a) shows the typical average I-V characteristics of the non-passivated MS device recorded by applying a voltage scan from -1000 V and +200 V on the pixelated electrodes. Despite the presence of a gold electrode on both sides, the as-fabricated CZT detector exhibits an asymmetric I-V characteristic, indicating different contact behavior at the metal-semiconductor interfaces. Zha et *al*. reported that Au deposited on CZT (111) A and (111) B surfaces, even when prepared by the same deposition method, yields different barrier heights owing to distinct interfacial chemical reactions [31]. Moreover, compared with the thermally evaporated Au contact, which typically forms sharp interfaces, the interface between CZT and electroless gold is heavily influenced by complex reactions during the electroless process, resulting in a chemically graded interface composed of Au/AuTe particulates and tellurium oxides [32,33]. In this work, the pixelated and planar electrodes were deposited by thermal evaporation and electroless processes, respectively, which may account for the rectifying behavior observed in the fabricated detectors [31,32,34]. The CZT samples used in the current study were indium-doped, which typically results in Fermi-level pinning at mid-gap or slightly n-type semiconductor behaviour [35]. A large barrier at the Au-CZT interface on the B-face prevents electron injection into the device when a negative bias is applied, and the detector operates in reverse bias, exhibiting a dark current of ~16 nA at -1000 V. Moreover, at a sweep delay of 1ms, the device shows a slightly higher current along with noticeable hysteresis in the I-V characteristic (Fig. 4(a)). This can be attributed to transient trapping/detrapping dynamics at the metal-semiconductor interface, especially at the electroless Au-CZT interface, as well as to shallow or deep-level traps within the CZT bulk[30,36,37]. Meanwhile, with a 500 ms source delay, the system reached steady state, and the hysteresis effect was largely suppressed.

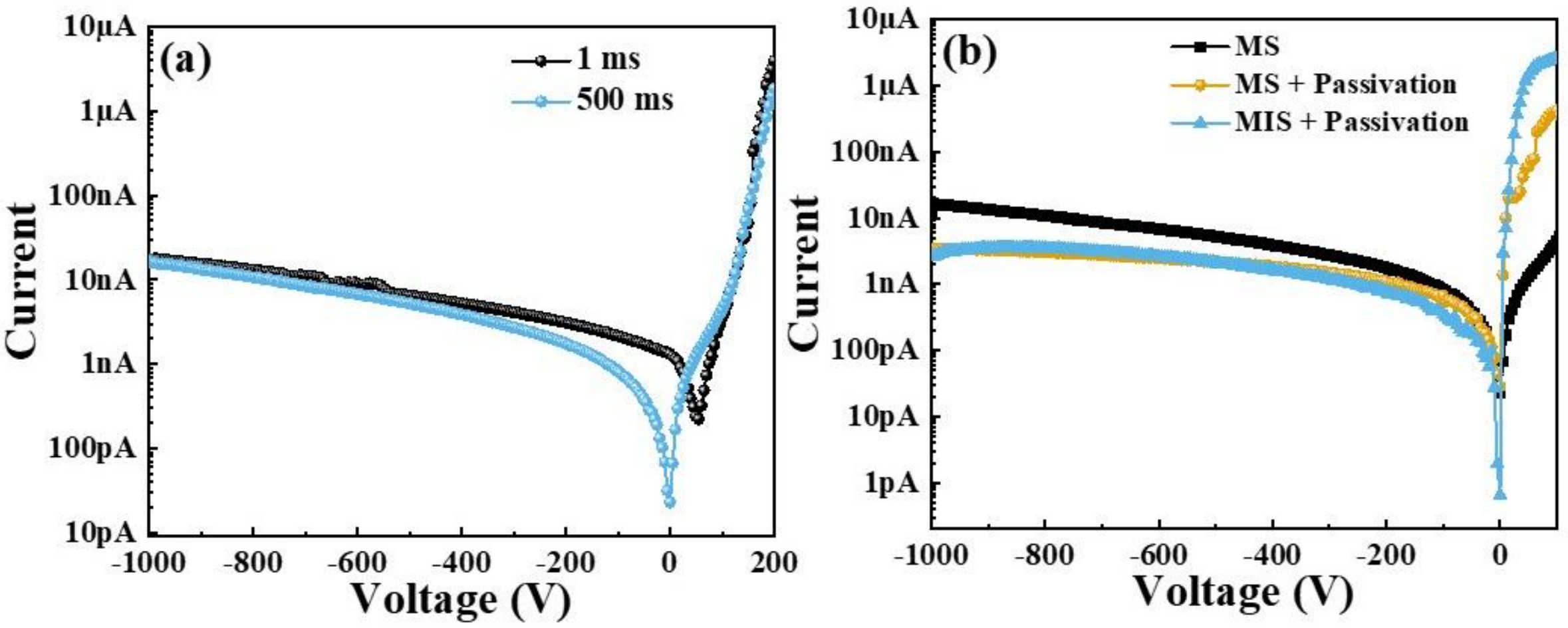

**FIG. 4:** (a) Typical I-V characteristics of the as-fabricated MS device recorded with two different source delay times under dark conditions. (b) Current-voltage characteristics of the MS, MS + passivation, and MIS + passivation devices measured in the dark.

The I-V characteristics of both MS and MIS devices were measured following $Al_2O_3$ passivation. Figure 4(b) shows the dark I-V characteristics of MS, MS + passivation, and MIS + passivation devices with a 500 ms source delay. It is evident that the dark current in both passivated devices is reduced by more than a factor of five at -1000 V, reaching values in the range of 2-3 nA. Although the dark current remains higher than the values reported by Bettelli *et al.* [38] for isolated pixels surrounded by guard-ring devices at similar electric fields, the detectors developed and investigated in this study exhibit a pixelated array layout without a guard-ring, which is more representative of a practical detector for imaging applications. Moreover, the rectification ratio of both MS and MIS devices increased after passivation, indicating that the thermal treatment during passivation likely improved the electroless Au-CZT interface. Notably, neither device exhibited hysteresis after passivation, as shown in Fig. S6, indicating that the traps were significantly annihilated at a relatively higher substrate temperature during the $Al_2O_3$ deposition process. Kim et *al*. reported that low-temperature annealing from 80℃ to 220℃ can form $TeO_2$ or $CdTeO_3$ beneath the electroless Au contact, leading to a reduction in dark current, but the annealing time in their case was much longer, exceeding 3000 minutes[39]. In contrast, the total time for $Al_2O_3$ deposition in the present work, including ramping up substrate temperature up to 250 °C, was approximately 45 minutes. Therefore, we attribute the observed reduction in dark current primarily to the effective passivation of the interpixel spacing and sidewalls of the CZT detectors, as supported by the decreased interpixel current measured before and after passivation. In addition, the passivation process conducted at a substrate temperature of 250 °C may also induce thermal activation at the electroless Au-CZT interface, potentially converting the Te-rich layer into $TeO_2$ or $CdTeO_3$. This thermally assisted interfacial modification may improve the quality of the metal-semiconductor contact, thereby further reducing the dark current. Nevertheless, under the present processing conditions, the elevated temperature does not appear to significantly affect the bulk properties of the CZT material. A slight fluctuation in the forward-bias dark current was observed in both devices after passivation, though the origin of this behaviour remains unclear.

In summary, we investigated the effect of high-temperature $Al_2O_3$ passivation on pixelated CZT detectors with two different device architectures, MS and MIS. ALD-grown $Al_2O_3$ deposited at 250 °C reduces the SLC by nearly an order of magnitude, lowering the interpixel dark current from 41 nA to ~2-3 nA at -200 V and decreasing the dark current by more than a factor of five at -1000 V between top-bottom electrodes. No significant difference in dark current was observed between the passivated MS and MIS devices. The observed reduction in leakage current is attributed to the suppression of surface states and defect-assisted charge-transport pathways after passivation. In addition, the high-temperature process likely introduces a mild annealing effect that mitigates trap states and improves the electroless Au-CZT interface, leading to the elimination of hysteresis and improved rectification behaviour in the I-V characteristics. The results demonstrate that high-temperature $Al_2O_3$ passivation is a promising strategy to improve the electrical performance and long-term stability of next-generation CZT-based radiation detectors.

See the Supplementary material for optical, SEM, IR transmission, and AFM images of CZT, XPS spectra of CZT, device processing and experimental details, AFM images, XPS spectra, and current-voltage characteristics of $Al_2O_3$ film, current-voltage characteristics of the MS and MIS device after passivation.

**Acknowledgement:**

The authors acknowledge the financial support of NSERC Canada, Canada Foundation for Innovation, and PRIMA Quebec.

**AUTHOR DECLARATIONS**

**Conflict of Interest**

The authors have no conflicts to disclose.

**DATA AVAILABILITY**

The data that support the findings of this study are available from the corresponding author upon reasonable request.

# Supplementary material:

## Reduced Dark Current in $Cd_{0.9}Zn_{0.1}Te$ Detector Arrays by Aluminium Oxide Passivation

Sudarshan Singh, Luc Montpetit, Eloïse Rahier, Mahmoud R. M. Atalla, Sebastian Koelling, Oussama Moutanabbir*

Department of Engineering Physics, École Polytechnique de Montréal, Montréal, C.P. 6079, Succ. Centre-Ville, Montréal, Québec, Canada H3C 3A7

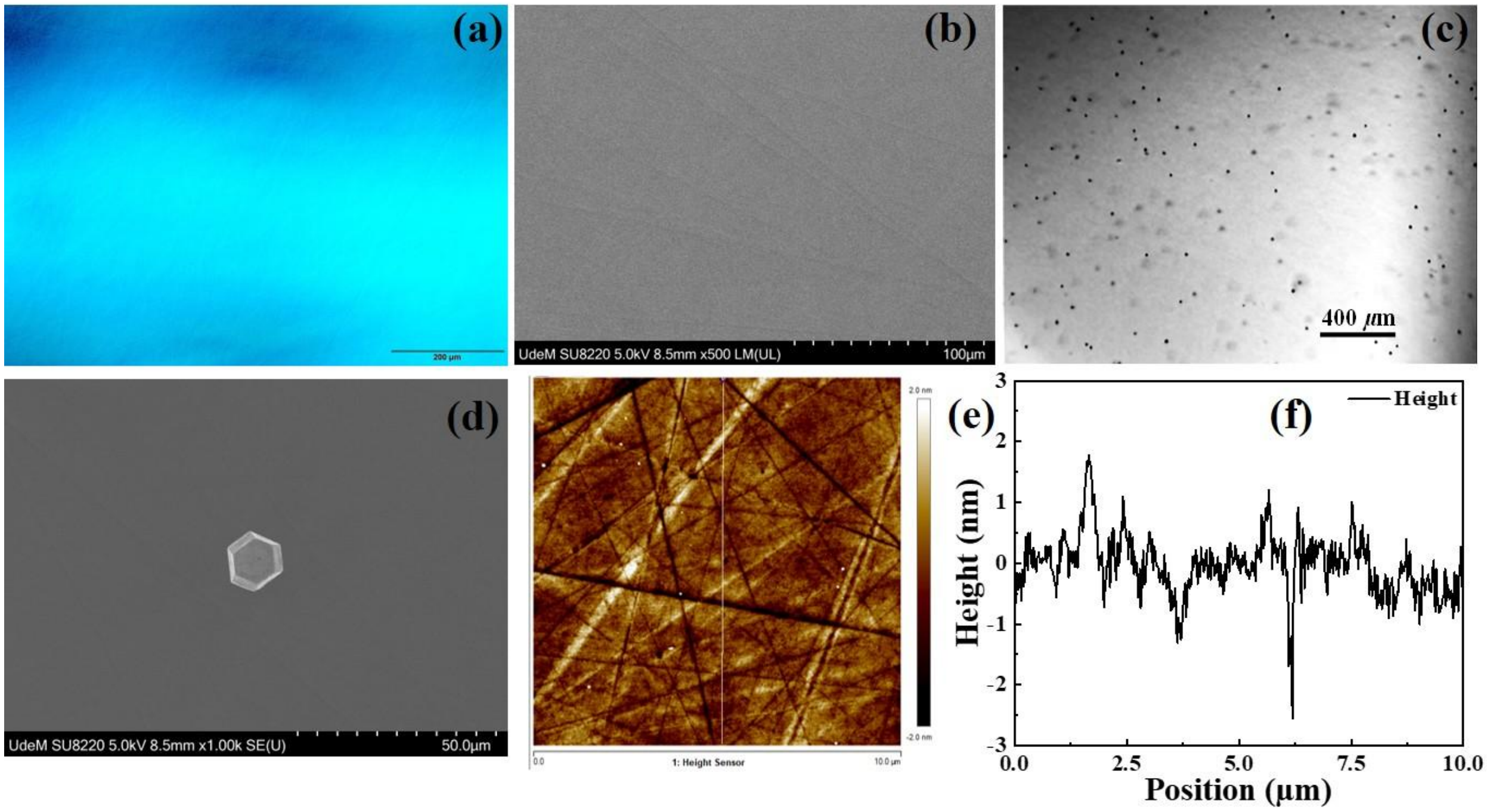

**FIG. S1:** (a) An optical microscope image of CMP-processed CZT samples after standard cleaning, and its (b) SEM and (c) IR images. (d) A typical top-view SEM image of the CZT sample showing a hexagonal-shaped Te-inclusion. (e) Typical AFM image representing the CZT surface and (f) line scan of the AFM image showing peak-to-valley height of polishing line.

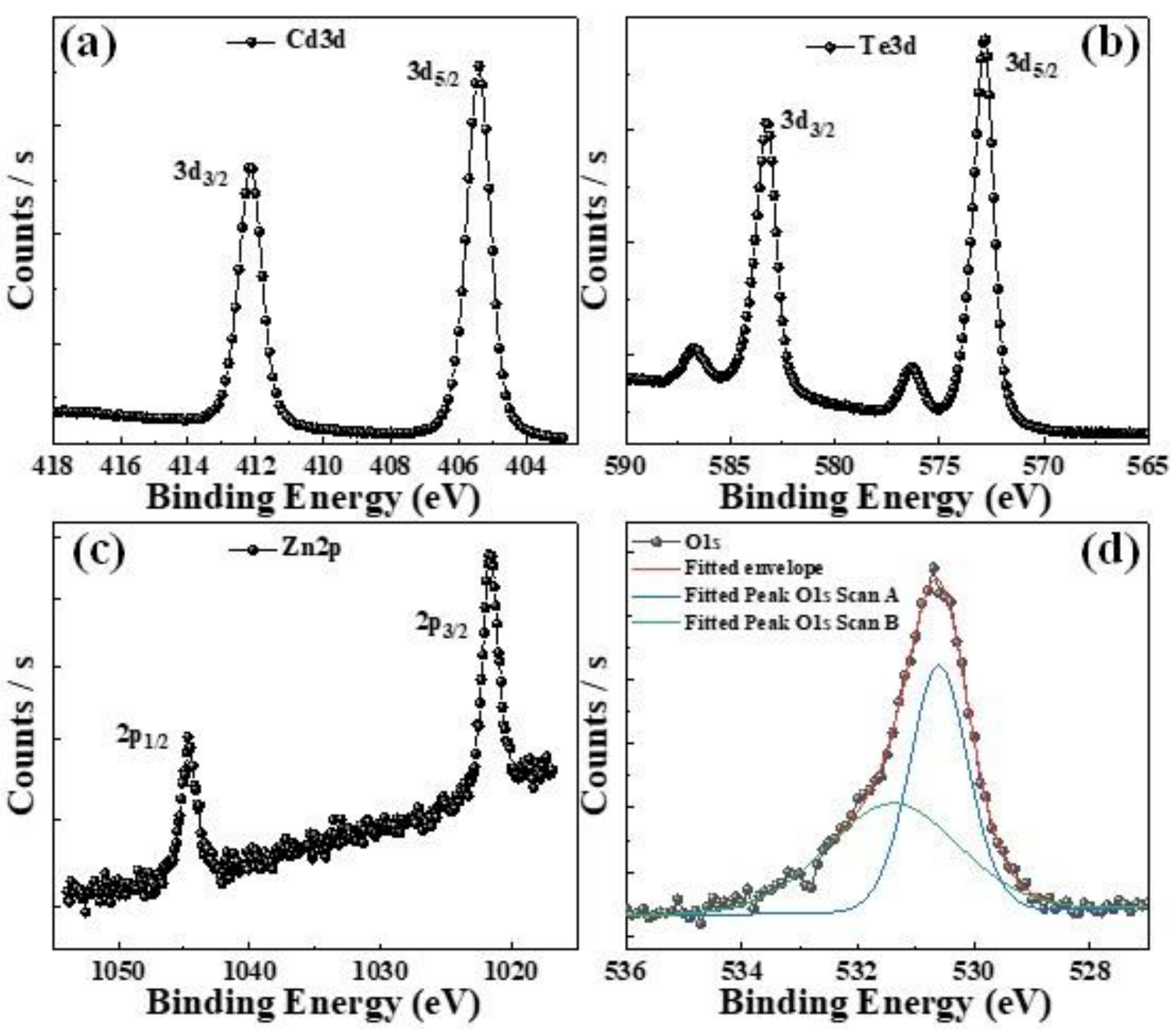

**FIG. S2:** High-resolution X-ray photoelectron spectra of (a) Cd-3d, (b) Te-3d, (c) Zn-2p and (d) O1s core electrons recorded from the CZT sample.

**Experimental details**

CZT ($Cd_{0.9}Zn_{0.1}Te$) wafers, oriented along the [111] direction, were chemo-mechanically polished. Thereafter, an electroless gold film was deposited on the A-face of a 2-inch CZT wafer using $AuCl_3$, and the wafer was then diced into several small pieces of size 10×10×2 $mm^3$. Before depositing the top electrode, the samples were sequentially cleaned via sonication for 5 minutes each in acetone, isopropyl alcohol (IPA), and deionized water, then dried with $N_2$ gas. The pixelated top electrodes, 500 × 500 $\mu m^2$ in size, were patterned on the B-face of the CZT samples using standard photolithography, followed by gold deposition via e-beam evaporation. The $Al_2O_3$ films were deposited using an atomic layer deposition (ALD) system at different substrate temperatures. High-purity water ($H_2O$) and trimethylaluminum (TMA) served as the oxygen and aluminium precursors, respectively. Each ALD cycle consisted of a 0.06 s dose pulse for both $H_2O$ and TMA, followed by a 4-second reaction time. Film thickness was calibrated based on the number of ALD cycles. Surface morphology was examined using an Olympus BX51M optical microscope. Infrared (IR) transmission images were captured with an infrared microscope equipped with a halogen light source and a solid-state IR camera. Scanning electron microscopy (SEM) images were recorded after sample cleaning. X-ray photoelectron spectroscopy equipped with Al Kα radiation (1486.6 eV) and a beam spot diameter of approximately 1 mm was used to analyze the surface composition of the samples. The samples were sequentially cleaned with acetone, isopropyl alcohol (IPA) and deionized water, then dried under $N_2$, and loaded into a high-vacuum chamber. Prior to measurement, the surface was further cleaned using an $Ar^+$ ion gun operated at two keV for 300 s to remove atmospheric contaminants. The sputtering parameters were carefully selected to preserve the integrity of the CZT surface. Transmission Electron Microscopy (TEM) lamellas were prepared with Focused Ion Beam (FIB) using standard lift-out procedures and final milling at 2 kV. TEM images (conventional or in scanning mode) were acquired using a 200 keV electron beam. Scanning TEM Energy-Dispersive X-ray Spectroscopy (EDS) was acquired using the same microscope, with an energy resolution of 127 eV. The I-V characteristics of the fabricated detectors were recorded using Keithley 4200 and 2470 instruments.

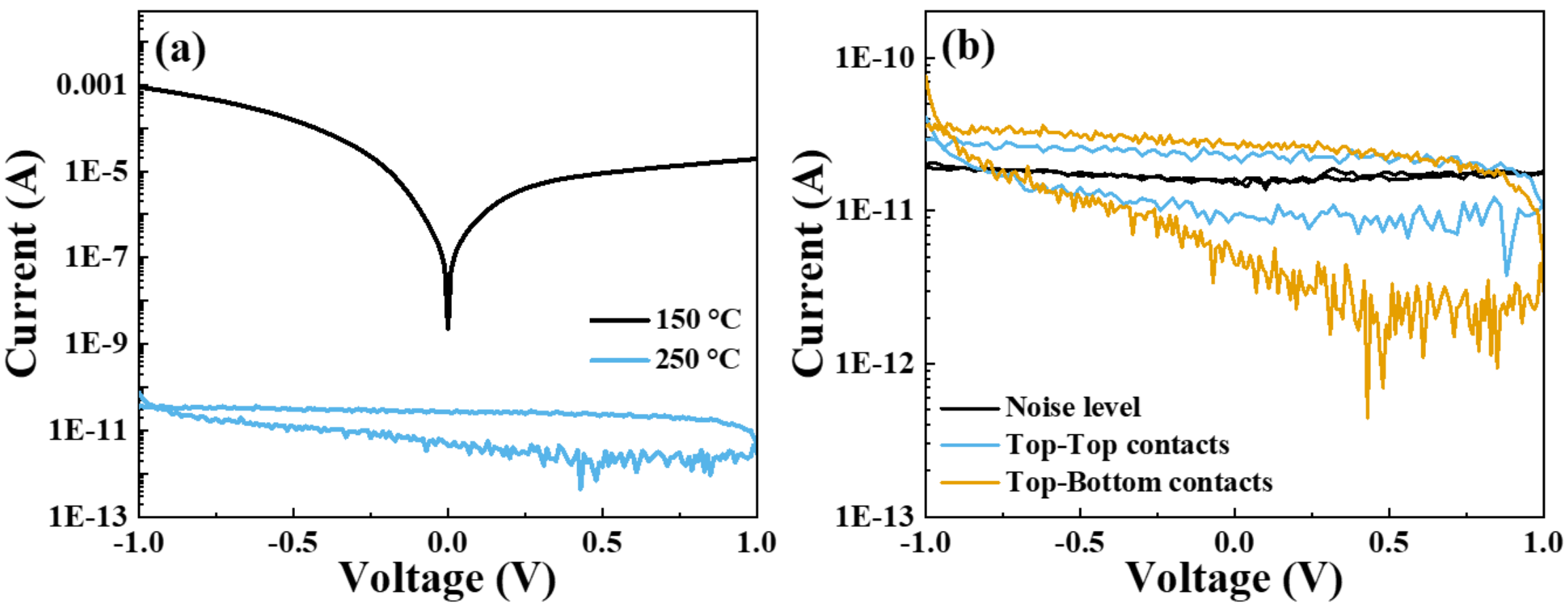

**FIG. S3:** (a) Typical current-voltage (I-V) characteristics measured across a 10 nm thick $Al_2O_3$ film deposited on Si at deposition temperatures of 150 °C and 250 °C (dual-voltage sweep). (b) I-V characteristics of the 10 nm thick $Al_2O_3$ film deposited at 250 °C, measured in different contact configurations. In top-top measurements, both electrodes were deposited on the $Al_2O_3$ layer; in top-bottom measurements, one electrode was deposited on the $Al_2O_3$ layer, and the other on the backside of the Si substrate. The noise level represents the current measured under open-circuit conditions.

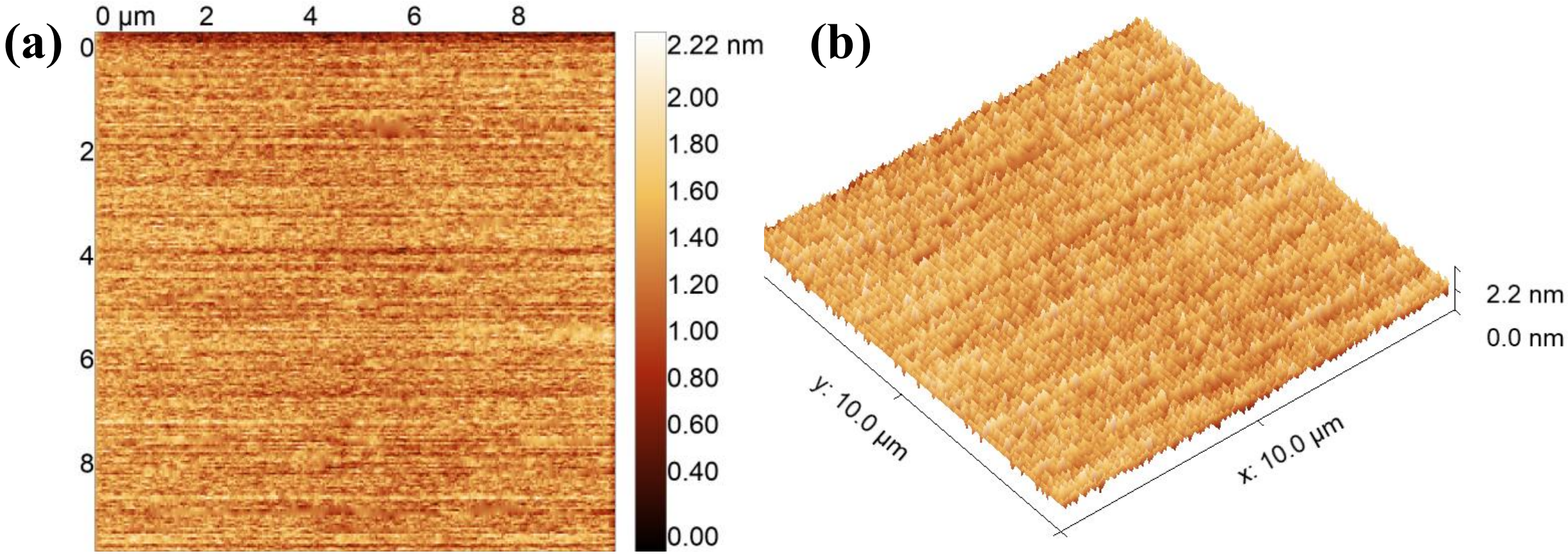

**FIG. S4:** Typical (a) 2D and (b) 3D AFM images of the 10 nm thick ALD-grown $Al_2O_3$ layer deposited on Si at a substrate temperature of 250 °C.

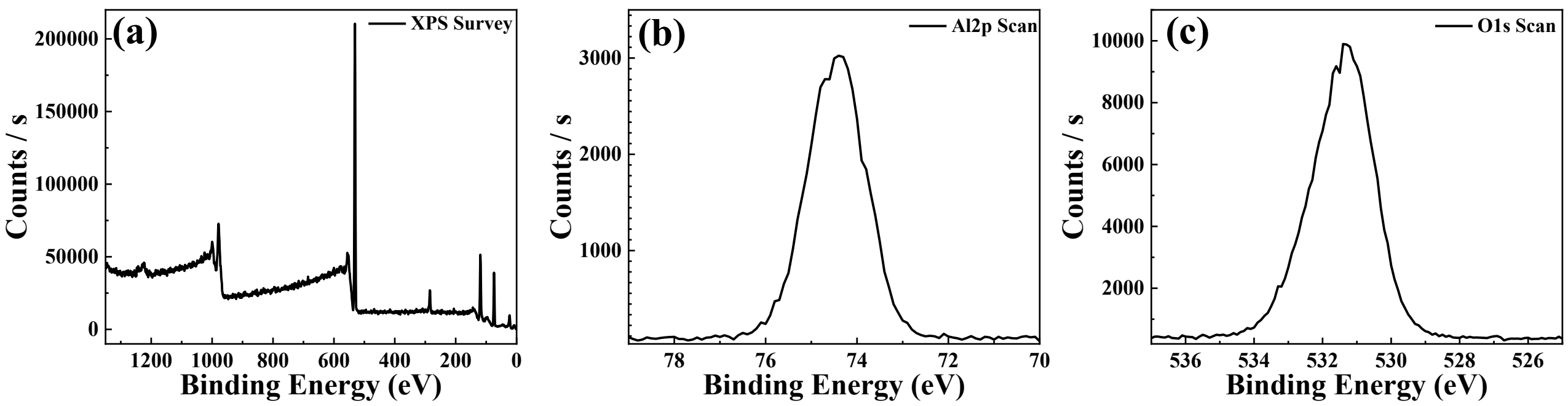

**FIG. S5:** X-ray photoelectron spectroscopy measurement illustrating the survey, Al2p, and O1s spectra of as-deposited $Al_2O_3$ layer of thickness 10 nm at a substrate temperature of 250 °C on a Si substrate.

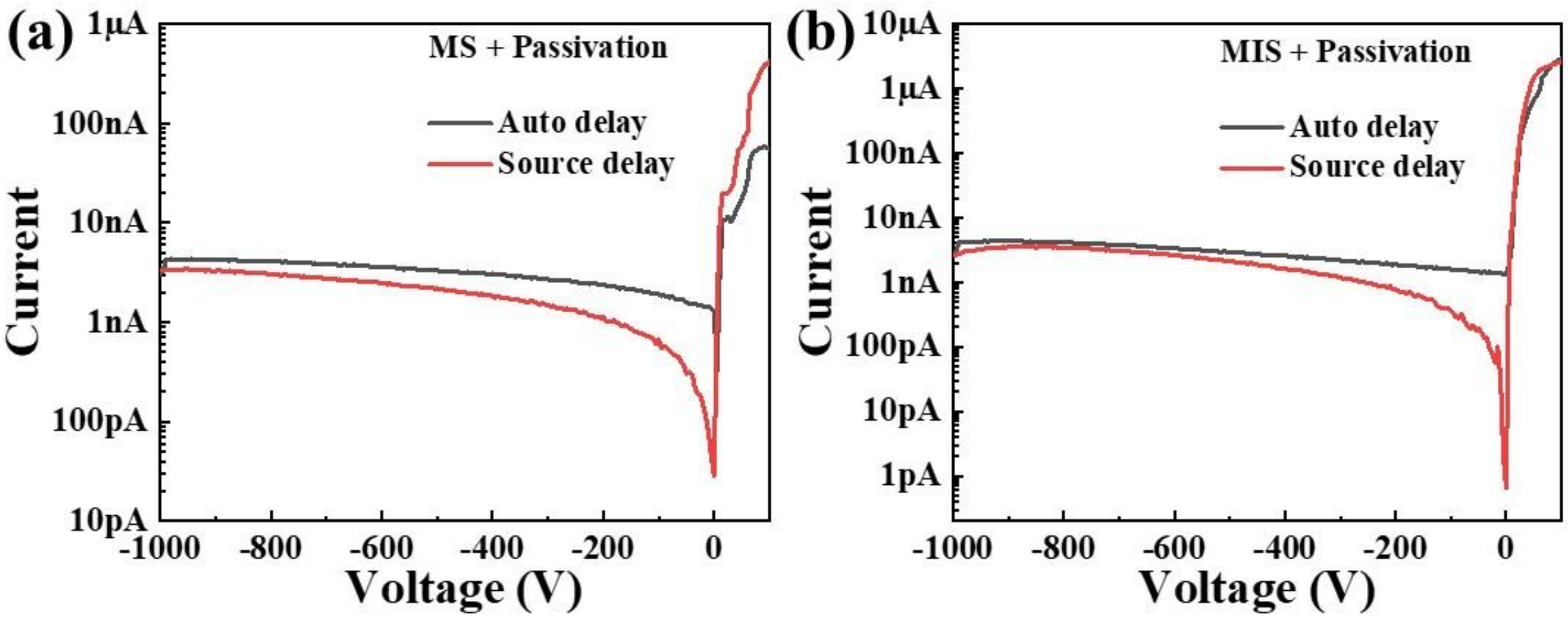

**FIG. S6:** Typical I-V characteristics of the (a) MS and (b) MIS devices after passivation, measured with auto delay (1ms) and source delay (500 ms) under dark conditions.